\documentstyle[aas2pp4]{article}
\tightenlines
\begin{document}
\def\ltsima{$\; \buildrel < \over \sim \;$}\
\def\simlt{\lower.5ex\hbox{\ltsima}}
\def\gtsima{$\; \buildrel > \over \sim \;$}
\def\simgt{\lower.5ex\hbox{\gtsima}}
\def\etal{et al.~}
\def\minspt{$^{\prime}_\cdot$}
\def\secspt{$^{\prime\prime}_\cdot$}

\title{Deep H-band Galaxy Counts and Half-light Radii from HST/NICMOS Parallel Observations}
\author{Lin Yan\altaffilmark{1}} 

\author{Patrick J. McCarthy\altaffilmark{1}, Lisa J. Storrie-Lombardi\altaffilmark{1}, \& Ray J. Weymann\altaffilmark{1}}

\altaffiltext{1}{The Observatories of the Carnegie Institution of Washington, \\
                  813 Santa Barbara St., Pasadena, CA 91101}

\begin {abstract}

    We present deep galaxy counts and half-light radii from F160W 
($\lambda_c=1.6\mu$) images
obtained with NICMOS on HST. Nearly 9 arcmin$^2$ have been
imaged with camera 3, with $3\sigma$ depths ranging from H = 24.3 to
25.5 in a 0.6$''$ diameter aperture.  
The slope of the counts fainter than H~$= 20$ is 0.31, and the
integrated surface density to H$\leq 24.75$ is $4 \times 10^5$ galaxies per
square degree. The half-light radii of the galaxies declines steeply
with apparent magnitude. At H~$=24$ we are limited by both
the delivered FWHM and the detection threshold of the images.
\end {abstract}

\noindent {\it Subject headings:} galaxies: evolution - cosmology: observations - 
infrared: galaxies.

\vskip 3cm
\centerline{\it Accepted for publication in ApJ Letters}

\section {Introduction}

Galaxy counts as a function of apparent magnitude probe both the
geometry of the Universe and the dynamical and luminosity evolution of
galaxies. Evolutionary effects dominate the departures of the counts from
the Euclidean expectation. The relative importance of the two primary forms
of evolution,
density and luminosity
evolution, can only be properly assessed with spectroscopic redshifts.
The near-IR pass-bands, however,
are better suited than visible colors
to purely photometric surveys as they are
less sensitive to 
star formation and extinction. The
weak dependence of
the k-correction on Hubble type (Poggianti 1997) and its slow change 
with redshift further enhance
the value of observing at wavelengths beyond $\sim 1\mu$m.

The deepest visible galaxy counts 
yield an integrated galaxy number 
density of 2$\times10^6/$ degree~$^2$ to m(F814W)$ = 29$ (Williams et al. 1996). 
The deepest K-band counts reach
K$ = 23.5$ with limited areal coverage ($\sim 2$~arcmin$^2$) and 
have a slope of $\rm d\log N/dm \sim 0.2-0.3$ for K$ > 18$
(Gardner, Cowie, \& Wainscoat 1993; Djorgovski et al. 1995; Moustakas et al. 1997).
The Near Infrared
Camera and Multi-Object Spectrograph (NICMOS; Thompson et al. 1998) on 
HST offers a means to extend the existing 
near-IR galaxy counts to much fainter levels in 
the equivalent of the H and J pass-bands. 
Parallel observations, in particular, provide
modest sky coverage for deep 
galaxy counts. At the depth of the 
deepest Keck/NIRC K ($2.2\mu$) counts our areal
coverage is presently $\sim 9$~arcmin$^2$ and will increase
as subsequent parallel exposures are taken. Our deepest
pointings presently cover an area comparable to the deepest
NIRC images, but are roughly one magnitude deeper.

This letter, and the independent work of Teplitz et al. (1998),
present the first deep H-band galaxy counts with NICMOS.
Our data reach H$\sim$24.8 (50\%\ completeness limit), 
deeper than all previous published near infrared
counts. The 
F110W ($\sim$ J) data, future parallel imaging
observations and a more detailed treatment of the present data
will presented in a separate publication.

\section{ Observations and Data Reduction}

The data were obtained with camera 3 of NICMOS
operating in the parallel mode, beginning in November 1997. 
The NICMOS internal pupil adjustment mirror was set
near the end of its range, providing the best possible focus for camera 3.
The PSF, measured from the stars in a selected globular cluster field,
is slightly non-gaussian, but is well characterized by
FWHM$ =
0.25^{''}$. 

We obtained 
four images per orbit, two each with the F110W ($\lambda_c=1.1\mu$) and F160W ($\lambda_c=1.6\mu$) filters.
The field offset mirror (FOM) was used to dither between two
fixed positions $1.8^{''}$ apart in a direction aligned with one axis of the detector.
In addition there were small inter-orbit dither moves executed for some of
the pointings. The detector was read using the STEP64 MultiAccum sample sequence,
with 13 or 18 samples, resulting in exposure times of 256 and 576 seconds,
respectively. The projected size of a camera 3 pixel
is $0.204{''}$, giving a $52.2{''} \times 52.2^{''}$ field of view for each image. A small area is lost in the construction of the final mosaic image.
We selected 12 fields with intermediate to high galactic latitudes
for this study (see Table 1).
The areal coverage of the present dataset is $8.7$~arcmin$^2$.

We used McLeod's (1998) NicRed v1.5 package to linearize and remove the cosmic
rays from the MultiAccum images.
Median images were derived from the 24 pointings and these were used
to remove the dark and sky signals.
Even with the optimal dark subtraction,
there remain considerable frame-to-frame variations in the quality of the
final images. We did not estimate photometric error for each individual
galaxy due to the frame-to-frame variation, however, as described below, 
our incompleteness simulation has taken into account this effect
for the galaxy number counts.
The individual linearized, dark corrected, flatfielded 
and cosmic ray cleaned images were
shifted, masked and combined to produce final mosaic images. Before shifting,
each image was $2 \times 2$ block-replicated and only integer ($0.1^{''}$) offsets
were applied. In this way we avoided any smoothing or interpolation of the data.
The MultiAccum 
process is not 100\% efficient in rejecting cosmic ray events and so we applied a
$3\sigma$ rejection when assembling the final mosaics.

\subsection{ Galaxy Detection \& Photometry}

We performed the object detection and
photometry using SExtractor version 1.2b10b 
(Bertin \&\ Arnouts 1996) and photometric zero points provided by 
M. Rieke (priv. comm) with an uncertainty of 0.05~mag.
Each image was convolved with
a gaussian kernel with FWHM $= 0.3^{''}$
for object detection, using a 2.0$\sigma$ detection threshold.
Isophotal magnitudes were measured to the $1\sigma$ isophot.

We define the
total magnitude in a manner similar to those used 
by Smail et al. (1995) and Djorgovski et al. (1995).
SExtractor calculates isophotal and aperture magnitudes for each galaxy, along with
the isophotal area at a given threshold. We adopted an aperture diameter
of $0.6^{''}$. For galaxies with isophotal diameters less than the 
$0.6^{''}$, we use their aperture magnitudes, after correction
to a 2$^{''}$ diameter aperture. In using the aperture magnitude
as a total magnitude, we assumed that all faint galaxies have 
similar profiles. We derived 
the aperture correction from an
average of 20 galaxies with isophotal diameters smaller than $0.6{''}$.
This average faint galaxy yielded an aperture
correction of 
$0.3\pm 0.05$ magnitudes, 
indistinguishable from that derived from stars with H$\sim 18$.

For galaxies with isophotal diameters between $0.6^{''}$ and
$2^{''}$, we measure isophotal magnitudes to 1$\sigma$ 
of sky and then correct to a $2^{''}$ diameter aperture
using a correction that is a function of the isophotal magnitude. 
In our deepest field, for H$<$22.5, the magnitude
correction is negligible, for 22.5$<$H$<$23.5 and
H$>$23.5, the corrections are 0.1 and 0.2 mags, respectively.
These corrections are derived for each field 
in the same fashion as the aperture correction
described above. 
Using the galactic foreground extinction A$_{\rm_B}$ derived
from the Burstein \&\ Heiles maps (Burstein \& Heiles 1984)
and the extinction law derived by Rieke \&\ Lebofsky (1985),
we find that the maximum extinction correction for our fields
is A$_{\rm H} \sim 0.03$ magnitudes.
The new dust maps
by Schlegel, Finkeiner \&\ Davis (1997) gives the maximum A$_{\rm H}$
of 0.05 mag. Thus, we did not apply any extinction correction to our photometry.

The sky variances, measured in 10 blank
$2.5^{''} \times 2.5^{''}$
areas in each image, give 1$\sigma$ surface brightness limits ranging from
24.8 to 26.0 mag/arcsec$^2$. This implies 3$\sigma$ limits
within a $0.3^{''}$ radius aperture of 24.3 to 25.5 magnitudes. 

\subsection{Incompleteness Modeling}

The raw counts must be corrected for false detections and incompleteness due to
crowding, flat-field and dark correction errors, and Poisson noise
before meaningful conclusions
can be drawn. 
We have carried out an extensive series of simulations to quantify 
the incompleteness effects.
We selected several well detected galaxies
from an image, dimmed them by various factors, and added these images into
the original image at {\em random locations}. We then apply the same
detection and photometry measuring algorithms as in the original
analyses. 
The use of random positions in the simulation
allows us to include completeness corrections arising from non-detections
and magnitude errors caused by crowding and spatially dependent errors in the sky subtraction
and flat field correction.
With this approach we create a matrix, $P_{ir}$, which is the probability
of finding a galaxy with the input magnitude of m$_i$ and the recovered magnitude m$_r$.
The number of input galaxies
$N_i$ at each magnitude m$_i$ was chosen to reproduce an initial guess for the 
slope of the counts.
The number of galaxies recovered at magnitude m$_i$, $N_r$, is
obtained from
$\sum_i P_{ir} \times N_i$. The detection rate $r_i$ is $N_r$/$N_i$, and the 
completeness correction at the corresponding magnitude bin is
1/$r_i$. For each image, we did more than $10^4$ simulations to estimate the
incompleteness corrections. 
We performed the incompleteness modeling for four
fields with representative integration times. The correction
was applied to the individual fields before the final 
galaxy number counts were produced. 

Of our 12 fields, 3 are at intermediate galactic latitudes. 
In these fields 
we identified stars using the SExtractor star/galaxy classifier.
We found 4 stars with magnitudes 20$<$H$<$22, 
which implies an average 2\%\ correction
to the number counts; for H$>$22.0, the correction at 
each magnitude bin is $<$2\%. At H$> 24$, 
the counts are dominated by compact galaxies in all of our fields. 
In addition, the measurements of half-light radii vs. total magnitude
suggest that the stellar contamination in all of our fields is
of the order of few percent. 

We estimated our false positive detection rate by reducing the 
``left'' and ``right'' dither
positions independently and by splitting our longest pointing into 
three independent, but shallower,
data sets. The ``left'' and ``right'' dither position images allowed 
us to assess the impact of
persistent cosmic rays, while the
photometric catalogs in the 1/3
exposure time images of 1120+1300 allowed us to assess the number of 
detections that had no corresponding object
in the deeper image. Both approaches yield false detection rates in the 
a few percent 
range in the faintest magnitude bin for each field. 

\subsection{ Half-light Radii}

We measured the half-light radius for each object in the final photometric
catalog. 
A subsection of the image centered on each object was extracted
and regrided with $20\times$ finer sampling. The enclosed flux was 
then computed
in $0.005^{''}$ steps (1/20 pixel) 
until one half of the total flux was enclosed. 
The code used for this
operation was kindly provided by I. Smail and so our results can 
be directly compared to
Smail et al. (1995). 
%The results for our 12 high latitude fields 
%are shown in Figure 1.
We measured the half-light radius of stars with
$16 < H < 23$ in one globular cluster field to derive the 
instrumental half-light
radius. The non-gaussian shape of the PSF requires this measurement 
to allow comparison
with the derived galaxy sizes. 

\subsection{ Results and Discussion}

Our deepest field, 1120+1300, is shown in Figure 1 (Plate []). We have circled
several representative faint galaxies and noted their magnitudes. The two brightest galaxies
in this field have isophotal magnitudes of 17.6 and 18.8, respectively. 
In Figure 2, we plot the raw and corrected differential galaxy number counts. 
Table 2 lists the raw counts, incompleteness corrected counts and
effective area at each 
magnitude bin. 
We determine a slope of $0.31\pm 0.02$
for  20$<$H$<$24.5, consistent with values derived at
K by various groups (Gardner et al. 1993; Djorgovski et al. 1995; Moustakas et al.
1997). We find no significant change in the slope for H $< 24.5$.
In the same figure, we plot the counts measured in I and K passbands 
by Smail et al. (1995), Williams et al. (1996), and Djorgovski et al. (1995). 
The integrated number of galaxies H$\leq 24.5$, including the incompleteness 
corrections, is $4 \times 10^5$ per sq. degree, or 2$\times 10^{10}$ galaxies
over the entire sky. This is about 3 times larger than 
the total implied from 
integration of the local luminosity function (Lin et al. 1996) to
0.01L$^\ast$ over an all-sky co-moving
volume for ($\Omega_0$, $\Omega_\Lambda$) $= (1, 0)$ cosmological model.

%\begin{figure}[]
%\figurenum{1}
%\plotone{newfig1b.ps}
%\caption{A greyscale representation of our deepest
%field, 1120+1300. Several faint objects are marked and their H magnitudes
%are noted. The field shown is $50.2^{''} \times 50.2^{''}$.}
%\end{figure}

\begin{figure}[]
\figurenum{2}
\plotone{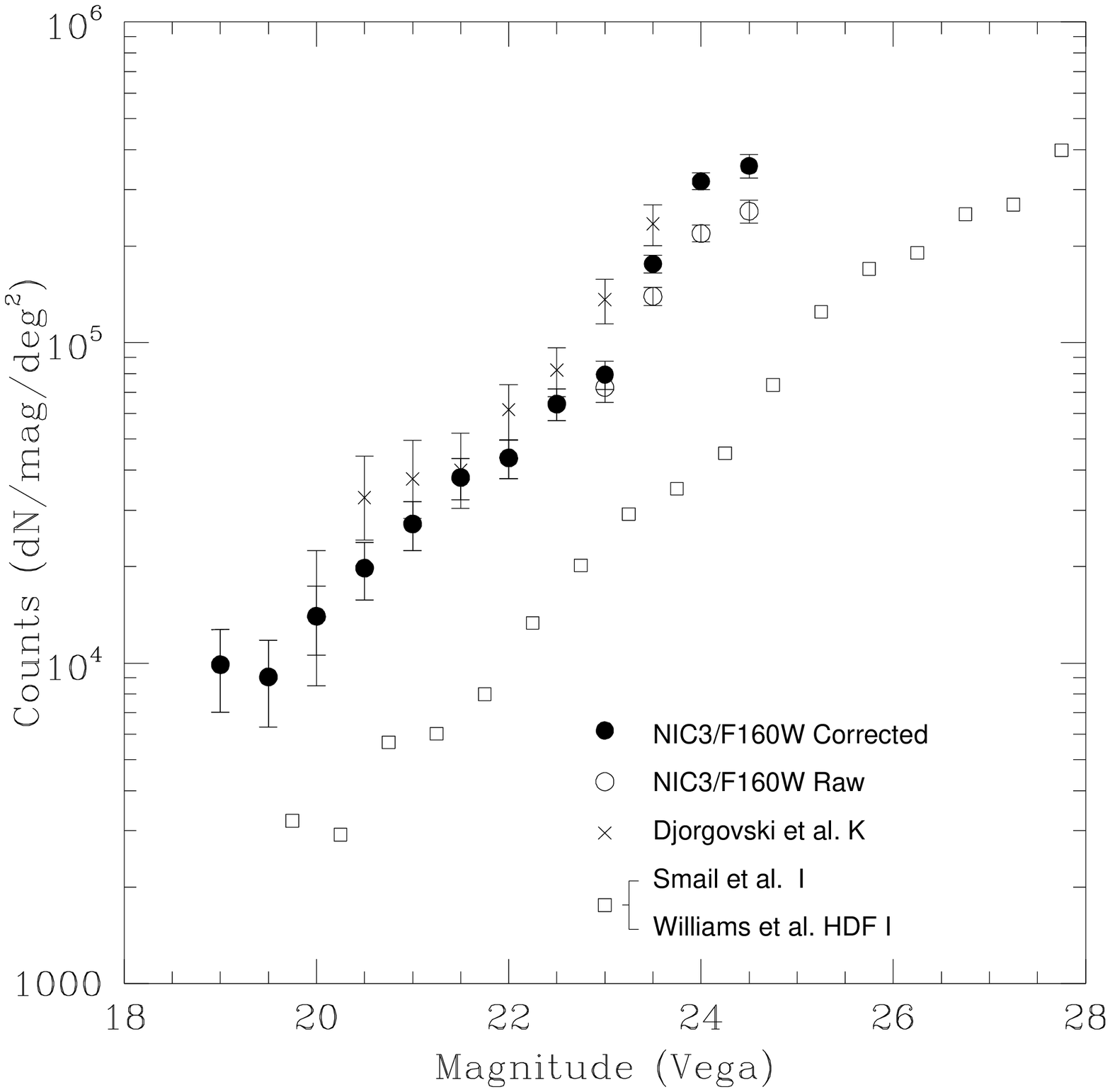}
\caption{The galaxy number count - magnitude
relation derived from our data and other surveys from the literature.
The open and filled circles are the raw and corrected NICMOS H counts,
respectively. The crosses are Djorgovski et al.'s K ($2.2\mu$) counts, the
open squares are the average of the I-band counts from Smail et al. (1995)
and Williams et al. (1996), with a zero-point shift of 0.48
magnitudes applied to the
AB magnitudes in Williams et al. to bring them to the system used in
Smail et al}
\end{figure}

%%%% Figure 2 - Counts
\begin{figure}[]
\figurenum{3}
\plotone{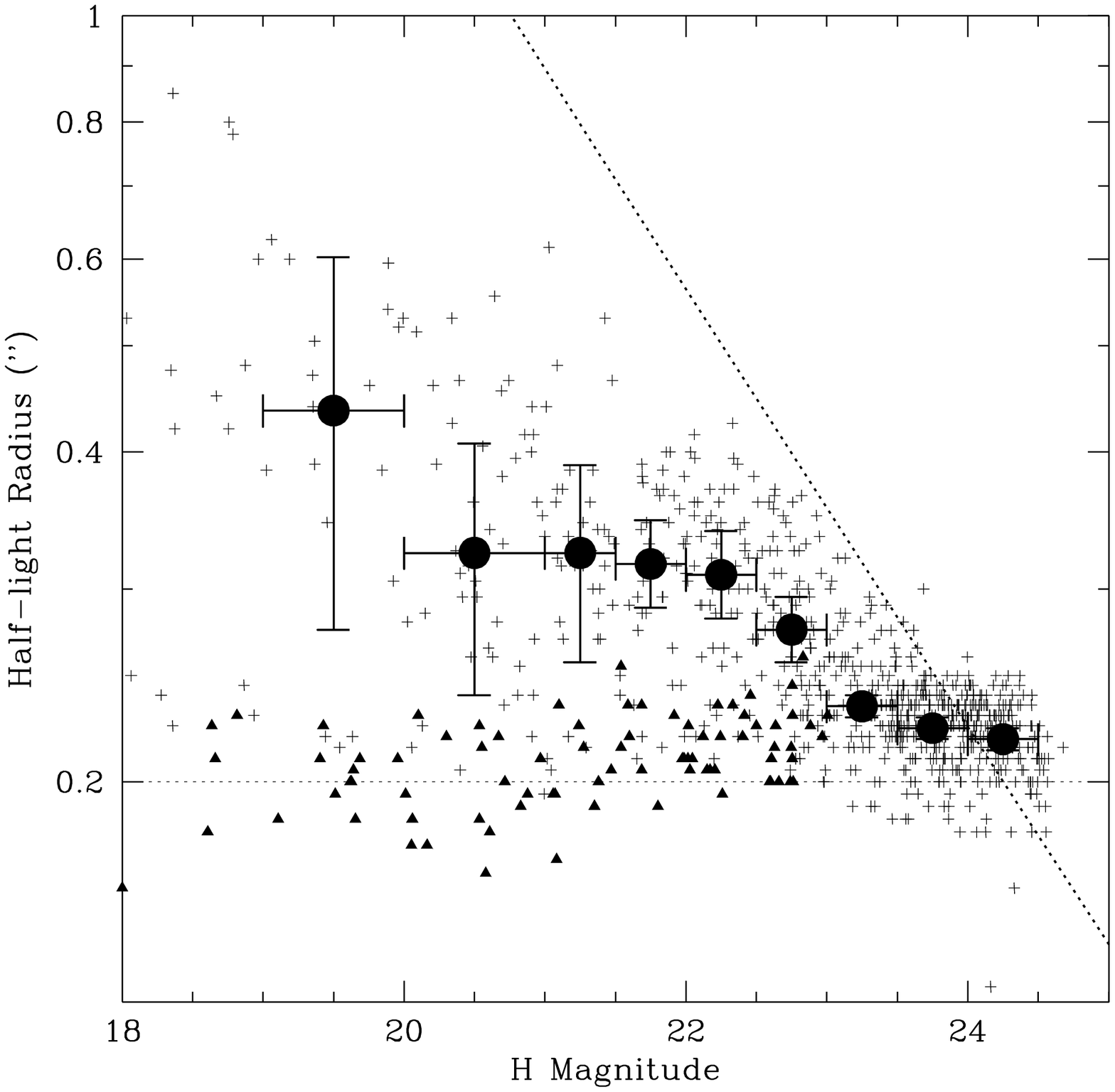}
\caption{
The half-light radii for all of the detected galaxies.
Stellar objects are indicated by star symbols and galaxies
by open circles. The solid dots indicate the median sizes in bins of 0.5
or 1 magnitude widths.
At H $=$ 24, we have reached the NIC3 angular resolution limit, 0.2''.
The diagonal dotted line shows the magnitude enclosed in an aperture of
radius $r$ with a uniform surface brightness of
22 magnitudes$/$arcsec$^2$.}
\end{figure}

The unambiguous interpretation
of the measured counts-magnitude relation in the
context of the galaxy formation and evolution requires
both knowledge of the redshift distribution and the faint-end 
slope of the luminosity function.
Our images probe the faint-end slope as well as the
distant universe. 
At z$=$1 H$=$24 corresponds to
a luminosity of $\sim \rm 0.01L^\ast$, 
(H$_0=$50 km$s^{-1}$ Mpc$^{-1}$, q$_0=$0.1, M$_K^\ast=$-24.75 (Gardner et al. 1997; Glazebrook et al. 1995)).
The K-band selected spectroscopic surveys of Cowie et al. (1996; 1997) reach a median
$z$ of unity at K$=$20-21.5, equivalent to H$\sim$21-22.5 for 
mild to passively evolving populations at z$\sim$1.
Thus our deep counts
must contain a large number of z$>$1 galaxies. 
The Lyman break selected galaxies
with spectroscopic redshifts also have K-band magnitudes 
of $\sim 21.5 - 23$ (Steidel et al. 1996),
within the depths of the counts presented here. 
What is unclear at present is if the
median redshift continues to climb at faint H and K magnitudes or 
if low luminosity
galaxies make an increasing contribution to the counts at H$>$23. 
This could be clarified by the photometric redshift measurements 
of these faint galaxies. 
The shallow and deep K-band surveys by Huang et al. (1997) and Cowie et al. 
(1996; 1997) find B$-$K colors that become increasingly
blue for galaxies with K$>$18, suggesting an increasing contribution from 
star forming systems. If the bulk of the galaxies that we detect at 
H $\sim 24$ are faint dwarf galaxies with 
luminosity much less than $\rm L^\ast$
at low redshifts instead of $\rm L^\ast$ galaxies at $z > 3$, 
the counts slope of $0.31$ 
implies a luminosity function $\Phi(L) \propto L^{-1.78}$, 
steeper than the local luminosity function. Although there is some evidence
that the faint end of the local luminosity function may be 
steeper than the standard value of $-1$ (Lilly et al. 1995), a slope of
$-1.78$ has not been observed. 

In Figure 3 we plot the measured 
half-light radii for all of the galaxies in our photometric 
catalog against their total magnitudes. 
We computed the median sizes in
bins of 0.5 or 1 magnitude
widths with the error bars as $\pm 1\sigma$ of the mean and these are shown as the solid symbols in Figure 3.
The dotted line corresponds to a uniform surface brightness of 
22 mag/arcsec$^2$.
It is clear from the crowding of
points at the faint end into the small area between the stellar locus and the
surface brightness limit, that there are strong surface brightness selection
effects in our sample at the faint end. Resolved objects with magnitudes 
fainter
than H$\sim$24 are undetectable in our data. For H$<$22 - 23, however, 
there appears to be a real deficit of galaxies near our surface brightness 
limit, suggesting that
the decrease in median half-light radius in the range $20 < H < 23$ is genuine. 
Images obtained with the high resolution cameras on NICMOS will 
allow one to extend the study of galaxy scale lengths to scales of 
$\simlt 0.1^{''}$, but the surface
brightness selection biases will, on average, be more severe for these data.

The NICMOS pure parallel program offers us
a wealth of unique data with high angular resolution and
unprecedented photometric depth in the H and J bands over an area of
$\sim 100$~arcmin$^2$. We will be able to carry out much more detailed 
studies, such as deep number counts
as functions of morphological types and colors, the evolution of 
galaxy intrinsic sizes. Combined with ground-based optical photometry
and spectroscopy, we may be able derive photometric redshifts for the 
numerous faint galaxies detected in the galaxy number counts. 

\section{ Acknowledgements}

We thank the staff of the Space Telescope Science Institute for 
their efforts in making this parallel program possible. In particular we
thank Peg Stanley and the staff of the PRESTO division, Bill Sparks, 
John Mackenty, and Daniella Calzetti of the NICMOS group and 
Bob Williams and Duccio Macchetto of the director's office.  
We acknowledge useful discussions with D. Hogg, I. Smail, 
R. Thompson, \& M. Rieke. B. McLeod and I. Smail are thanked for generously 
allowing the use of their software for the data reduction and scale length
measurements. This research was supported, in part, by grants
from the Space Telescope Science Institute, GO-7498.01-96A and 
P423101.
This project made use of the NASA/IPAC Extragalactic Database, 
operated by the Jet Propulsion Laboratory, California Institute of 
Technology, under contract with the National Aeronautics and 
Space Administration.

%\vfill
\bigskip
\bigskip
%\eject

%\eject

\onecolumn
\begin{table}
\centering
{\bf Table 1. The Selected Fields}
\begin{tabular}{cccccccccc}\\ \hline \noalign{\medskip}
Field & RA & DEC & l & b & T(F160W) &  $\mu(1\sigma)$ & $m_{lim}$ & $N_{lim}$\\
      & \multispan{2}\hfil (J2000) \hfil & (deg) & (deg) & (sec) & mag/$\Box ''$& $50\%$ & $50\%$ \\ \hline\hline
0240$-$0141 & 02:40:13 & -01:41:27 & 173 & -54 & 2816  &  24.8 & 23.8 &  62 \\
0304$-$0015 & 03:04:38 & -00:15:04 & 178 & -48 & 2560  &  25.3 & 23.8 &  25 \\
0457$-$0456 & 04:57:19 & -04:56:51 & 204 & -28 & 4480  &  25.7 & 24.3 &  42 \\
0729$+$6915 & 07:29:57 &  69:15:02 & 146 &  29 & 5120  &  25.9 & 24.5 &  31 \\
0744$+$3757 & 07:44:32 &  37:57:21 & 182 &  26 & 2048  &  25.1 & 23.5 &  49 \\
1039$+$4144 & 10:39:37 &  41:44:54 & 176 &  59 & 5376  &  25.5 & 24.5 & 74 \\
1120$+$1300 & 11:20:35 &  13:00:00 & 242 &  65 & 13824 &  26.0 & 24.8 & 137 \\
1237$+$6215 & 12:37:33 &  62:15:27 & 126 &  55 & 4608  &  25.7 & 24.3 &  47  \\
1604$+$4318 & 16:04:55 &  43:18:56 &  69 &  48 & 3840  &  25.6 & 24.0 &  53  \\
1631$+$3001 & 16:31:39 &  30:01:23 &  50 &  42 & 3584  &  25.1 & 24.0 &  88  \\
2220$-$2442 & 22:20:12 & -24:42:00 &  28 & -56 & 5120  &  25.9 & 24.5 & 71 \\
2344$-$1524 & 23:44:00 & -15:24:50 &  66 & -70 & 3840  &  25.4 & 24.0 & 60 \\ 
\hline
\end{tabular}
\end{table} 

%\onecolumn
\begin{table}
\centering
{\bf Table 2. Raw and Corrected Counts} \\
\begin{tabular}{cccccc}\\ \hline \noalign{\medskip}
Mag. & $\rm \log N_{raw}$ & $\rm \log N_{corr}$ & $-\sigma$ & $+\sigma$ & A$_{eff}$ \\ 
 &  deg$^{-2}$mag$^{-1}$ & deg$^{-2}$mag$^{-1}$ & & & arcmin$^2$ \\ \hline\hline
19.0 & 3.995 & 3.995 & 0.148 & 0.11 & 8.7  \\
19.5 & 3.957 &  3.957 & 0.156 &  0.114 & 8.7  \\
20.0 & 4.146 & 4.146 &  0.121&  0.094 & 8.7 \\
20.5 &  4.296  &    4.296 &   0.099 &  0.081 & 8.7  \\ 
21.0 &  4.434  &   4.434 &   0.083  &0.070  & 8.7  \\
21.5 &   4.579 &   4.579 &   0.069  & 0.0597 & 8.7  \\
22.0 &   4.640 &   4.640 &   0.064  &0.056  & 8.7  \\
22.5 &  4.808  &   4.808 &     0.052 & 0.0466  & 8.7  \\
23.0 &   4.860 & 4.900 &   0.047 &  0.042   & 8.7  \\
23.5 &  5.144 &  5.267 &   0.028 & 0.028  & 8.7  \\
24.0 &  5.340 &  5.503 &   0.026 & 0.026   & 6.5  \\
24.5 &  5.410 &  5.551 &    0.038 & 0.035    & 2.9  \\
\hline
\end{tabular} \\ {\medskip}
Note: Errors are in logarithmic scale and A$_{eff}$ is the effective\\
\ \ \ \ \ \ \  area at each magnitude bin. \ \ \ \ \ \ \ \ \ \ \ \ \ \ \ \ \ \ \ \ \ \ \ \ \ \ \ \ \ \ \ \ \ \ \ \ \ \ \ \ \ \ \ \ \ \ \ \ \ \ \ 
\end{table}

\end{document}